\documentclass[letterpaper, 10 pt, conference]{ieeeconf}  

\IEEEoverridecommandlockouts                              

\usepackage{verbatim}
\usepackage{cite}
\usepackage{amsmath,amssymb,amsfonts}
\usepackage{algorithmic}
\usepackage{graphicx}
\usepackage{textcomp}
\usepackage{hyperref}
\usepackage{caption}
\usepackage{subcaption}
\usepackage{siunitx}
\usepackage{flushend} 
\usepackage[monochrome]{xcolor}
\def\BibTeX{{\rm B\kern-.05em{\sc i\kern-.025em b}\kern-.08em
    T\kern-.1667em\lower.7ex\hbox{E}\kern-.125emX}}

\begin{document}
\title{The LuViRA Dataset: Synchronized Vision, Radio, and Audio Sensors for Indoor Localization}
\author{Ilayda Yaman$^{1}$, Guoda Tian$^{1}$, Martin Larsson$^{2}$, Patrik Persson$^{2}$, Michiel Sandra$^{1}$, Alexander D\"urr$^{3}$,   \\ 
Erik Tegler$^{2}$, Nikhil Challa$^{1}$, Henrik Garde$^{4}$, Fredrik Tufvesson$^{1}$, Kalle \r{A}str\"om$^{2}$, Ove Edfors$^{1}$,  \\
Steffen Malkowsky$^{1}$, and Liang Liu$^{1}$
\thanks{This work is funded by the Swedish Research Council, Ericsson AB, and ELLIIT (Excellence Center at Linköping-Lund in Information Technology).}%
\thanks{The authors are with Lund University, $^{1}$Department of Electrical and Information Technology, $^{2}$Centre for Mathematical Sciences, $^{3}$Department of Computer Science, $^{4}$Humanities Lab (Corresponding author email:
{\tt\small ilayda.yaman@eit.lth.se})}%
}

\maketitle

\begin{abstract}
 We present a synchronized multisensory dataset for accurate and robust indoor localization: the Lund University Vision, Radio, and Audio (LuViRA) Dataset. The dataset includes color images, corresponding depth maps, inertial measurement unit (IMU) readings, channel response between a 5G massive multiple-input and multiple-output (MIMO) testbed and user equipment, audio recorded by 12 microphones, and accurate six degrees of freedom (6DOF) pose ground truth of 0.5 mm. We synchronize these sensors to ensure that all data is recorded simultaneously. A camera, speaker, and transmit antenna are placed on top of a slowly moving service robot, and 89 trajectories are recorded. Each trajectory includes 20 to 50 seconds of recorded sensor data and ground truth labels. Data from different sensors can be used separately or jointly to perform localization tasks, and data from the motion capture (mocap) system is used to verify the results obtained by the localization algorithms. The main aim of this dataset is to enable research on sensor fusion with the most commonly used sensors for localization tasks. Moreover, the full dataset or some parts of it can also be used for other research areas such as channel estimation, image classification, etc.
 Our dataset is available at: \href{https://github.com/ilaydayaman/LuViRA_Dataset}{https://github.com/ilaydayaman/LuViRA\_Dataset}
\end{abstract}


\section{Introduction}
Fully autonomous driving and human-free smart factories are expected to significantly improve life quality~\cite{WANG2016158, 9046805}. To build an autonomous smart factory, one of the most critical challenges is performing accurate localization and monitoring of autonomous service robots in real-time for such an indoor environment, e.g., a service robot in a factory needs to localize itself within centimeter-level accuracy to perform tasks such as lifting and placing objects. To solve this problem, localization algorithms that use multiple sensors to improve accuracy and reliability have developed immensely~\cite{ICRA_multi_sensor, sensor_fusion_wifi, ICRA_sensor, radio_fusion}. 
There are many works that use cameras, radio receivers and transmitters, microphones, and speakers separately to localize objects in indoor environments~\cite{DROID-SLAM, MIMO_loc, zhayida2014automatic}. Recent works~\cite{radio-visual, Audio_Visual} indicate that these sensors can be used jointly to complement each other and enhance the overall performance. However, to develop, evaluate, and compare algorithms that fuse these sensors, a public dataset that includes simultaneous and synchronized sensor readings from each sensor in the same environment is required. 

Public datasets and benchmarks are key to the evaluation and development of algorithms designed by the scientific community. 12-Scenes~\cite{12scenes}, KU Leuven~\cite{mimo_dataset} and Stucture from Sound~\cite{StructureFromSound} datasets are examples of public datasets that can be used for vision, radio, and audio based localization algorithms. 12-Scenes is one of many indoor localization datasets based on vision sensors~\cite{survey_slam}, but very few public datasets include radio and audio sensors. 
Massive multiple-input and multiple-output (MIMO) is an enabling technology for the fifth-generation technology standard for cellular networks, or 5G, which has a scarcity of comprehensive datasets, especially for localization tasks where ground truth labels are required. The KU Leuven dataset is one of the very few datasets that contains measured channel state information (CSI) between a user and a massive MIMO testbed. The StuctureFromSound dataset includes audio data collected with $12$ microphones. These existing datasets contain the target sensors separately. Fusing separate datasets together for multisensory applications is not possible due to the lack of synchronization of different sensors as well as the different environments where the data has been captured. 

\begin{figure}[tb!]
  \centering
  \includegraphics[width=1.0\linewidth]{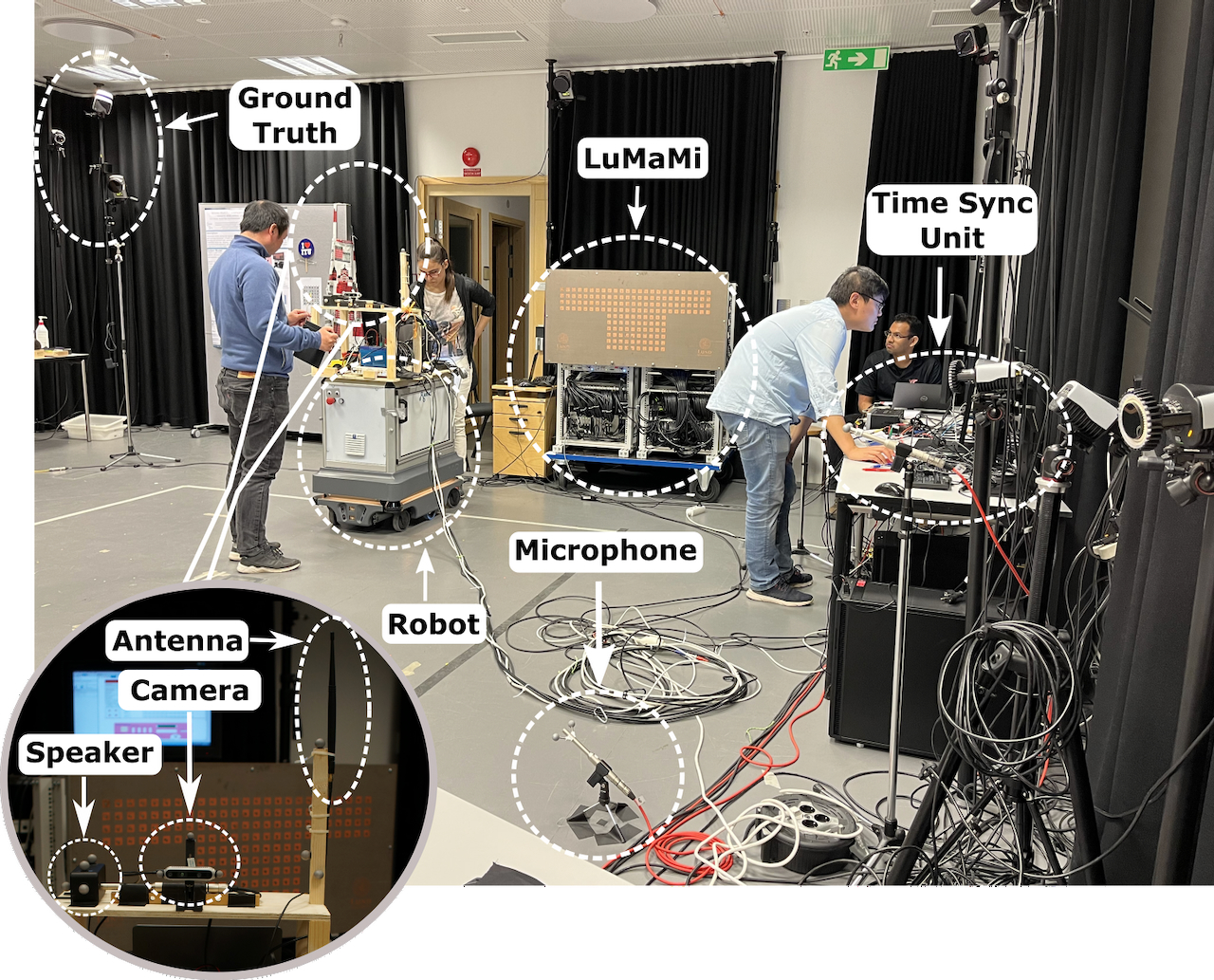}
  \caption{A photo of the sensors and the ground truth system depicting labels for the different components and the bottom-left corner focuses on the devices on the robot.}
  \label{fig:high_level_diagram}
\end{figure} 

In this paper, we describe, to the best of our knowledge, the first publicly available dataset that includes synchronized data from vision, 5G radio, and audio sensors that are captured in the same indoor environment. We name the dataset that consists of $89$ trajectories \textbf{the Lund University Vision, Radio, and Audio Dataset} or \textbf{LuViRA}. 
A \textit{Mobile Industrialized Robot} (MIR200) is used to carry the required equipment for all the aforementioned sensors in the Lund University Humanities Lab's Motion Capture (Mocap) Studio. The radio data is recorded using the world's first real-time 5G massive MIMO testbed, \textit{LuMaMi}~\cite{LuMaMi}. A photo of the Mocap Studio, including the sensors, and the ground truth system, is given in Fig.~\ref{fig:high_level_diagram} where each system and other required equipment are labeled. The detailed measurement setup, properties of each sensor, the synchronization schemes, key results, and known limitations of the dataset are described in the paper. The dataset is verified using state-of-the-art localization algorithms for the different sensors.   


\section{Measurement Setup} 

Ground truth systems, that achieve higher localization accuracy than the state-of-the-art algorithms and use the target sensors, are crucial for the precise validation and benchmarking of these algorithms, making them an important component in our measurement campaign.
The Lund University Humanities Lab's Mocap Studio consists of $18$ networked high-speed infrared \textit{Qualisys} cameras that record at $100$\,Hz and are connected to the \textit{Qualisys Sync Unit}, a synchronization box, for accurate synchronization. 
The system is calibrated and the localization error is measured to be less than $0.5$\,mm in the effective area. Due to the high accuracy that is achieved, it is used as a ground truth system throughout all the measurements.

\begin{figure}[tb!]
  \centering
   \includegraphics[width=0.90\linewidth]{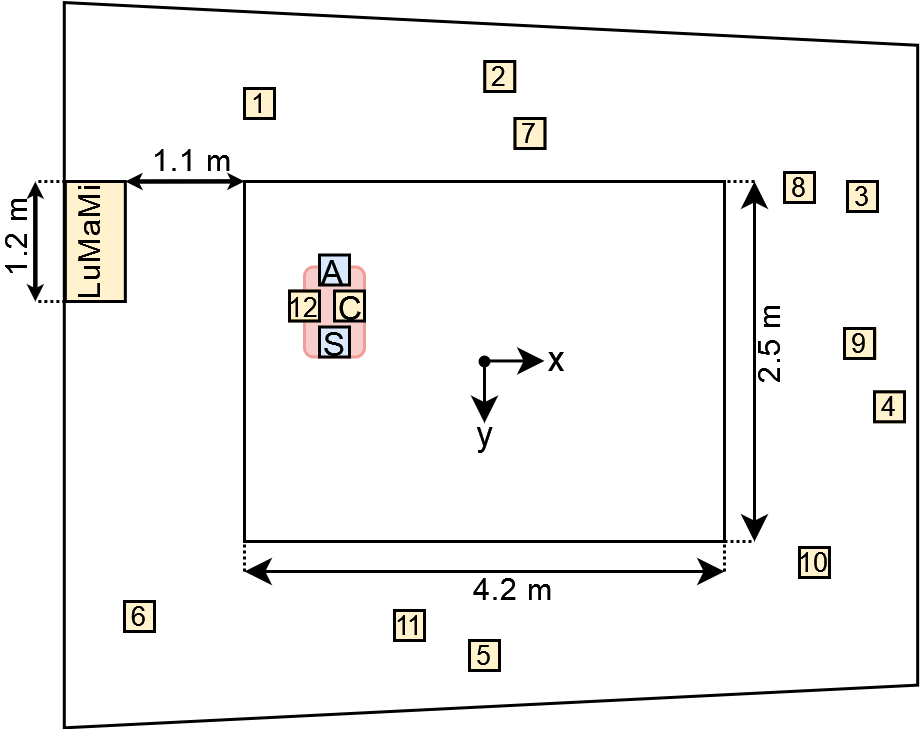}
  \caption{Top-down view of the studio, containing twelve microphones (1-12), antenna (A), RGB-D camera (C), speaker (S), and LuMaMi testbed.}
  \label{fig:main_view}
 \end{figure}

A $4.2 \times 2.5 \textrm{\,m}^{2} $ area shown in Fig.~\ref{fig:main_view} is chosen as the effective area where the trajectories are recorded. The size of the effective area is decided on the basis of many factors, but the main reason is the coverage of the ground truth system. The ground truth system tracks the 3D positions of the robot and the sensors by placing reflective spherical markers on target objects. 
Moreover, with a rigid setup of at least $3$ markers on an object, the system provides six degrees of freedom (6DoF) data, i.e., 3D orientation plus the 3D position. 
The ground truth system is the most accurate when several cameras can see the markers and on the sides of the studio, some markers are usually blocked by other objects.

Many precautions are taken to create a controlled environment for all the systems in our measurement campaign. A mobile industrial robot is used to securely transport a camera, a single antenna user equipment (UE), a sound source (speaker), and the equipment connected to them. The camera, antenna, and speaker are placed on a higher-level platform built for this measurement campaign. The platform ensures that the sensors have dedicated and stable locations on the robot and have an unobstructed view of the environment. A top-down view of the sensor placement in the studio is shown in Fig.~\ref{fig:main_view}, with sensors marked in yellow and actuators marked in blue. The rounded red rectangle depicts the robot, squares marked with ``A", ``C" and ``S" stand for antenna, camera, and speaker, respectively. The \textit{LuMaMi} testbed is placed in the top-left corner, and the placement of the microphones is illustrated as numbered squares. 

\section{Sensor Setup} 

This section gives detailed descriptions of the vision, radio, and audio systems. A summary of all the sensors recorded and their characteristics are shown in Table \ref{tab:table_compare}. For completeness, the ground truth system is also added to the table. As can be seen in Table \ref{tab:table_compare}, different sensors have different snapshot rates per second (ps). Thus, when these sensors are used together, the data should be associated with each other based on their respective timestamps. An example code that uses the sensor with the lowest snapshot rate as a reference and maps the others using the smallest time difference between timestamps is provided with the dataset.

\begin{table}[htbp]
\caption{Overview of the four main systems.}
\begin{center}
\begin{tabular}{|l|c|c|c|}
     \hline
     \textbf{System} & \textbf{Snapshot} & \textbf{Sensor}  &  \textbf{File} \\ 
     & \textbf{Rate (ps)} & & \textbf{Format}  \\ \hline
     Vision & $15$-$30$ & Intel® RealSense™ & .png and\\ 
      & \,frames & depth camera  & .bag \\\hline
     Radio & $100$ channel & LuMaMi testbed & .mat \\
      & estimates & with $100$ antennas & \\\hline 
     Audio & up to $96$\,k& 12 & .wav \\ 
      & samples & microphones & \\\hline
     Ground & $100$ & 18 high & .txt \\ 
     Truth & frames & speed cameras  &  \\\hline
  \end{tabular}
  \label{tab:table_compare}
  \end{center}
\end{table}

\subsection{Vision system}
An \textit{Intel® RealSense™ D435i} camera connected to a laptop on the robot is used as the input for the vision system. Each snapshot of visual data includes an RGB or color image, left and right infrared images (
obtained by imagers), a depth map, a point cloud, and inertial measurement unit (IMU) data. The IMU data is collected by two sensors: an accelerometer and a gyroscope. 
The features are captured by using Robot Operating System (ROS) and rosbag (.bag) file format. A summary of all the features of the camera is given in Table \ref{tab:table_camera}. The frequency of different features is chosen on the basis of the maximum bandwidth of the USB-2 bus on the laptop. For future dataset creation, we instead suggest choosing a common fps for the RGB, infra cameras and depth map, as well as a common frequency for accel, and gyro. This would simplify usage and a reduced fps would potentially allow higher resolution. 

\begin{table}[htbp]
\caption{Features and Recording settings of the Camera}
\begin{center}
\begin{tabular}{|l|c|c|c|}
     \hline
     \textbf{Sensor} & \textbf{Snapshot} & \textbf{Resolution} &  \textbf{File} \\ 
     & \textbf{Rate} &  &  \textbf{Format}  \\ \hline
     RGB Camera & $30$\,fps & $640$x$480$ & .png and .bag \\\hline 
     Left Imager & $15$\,fps & $640$x$480$ & .bag \\\hline
     Right Imager & $15$\,fps & $640$x$480$ & .bag \\\hline
     Depth Module & $15$\,fps & $640$x$480$ & .png and .bag \\\hline
     IMU (gyroscope) & $400$\,Hz  & N/A & .bag \\\hline
     IMU (accelerometer) & $100$\,Hz  & N/A & .bag \\\hline
  \end{tabular}
  \label{tab:table_camera}
  \end{center}
\end{table}

\begin{figure}[tb!]
  \setlength{\fboxrule}{0pt}
  \framebox{\parbox{3in}{
  \centering
  \includegraphics[height=0.70\linewidth]{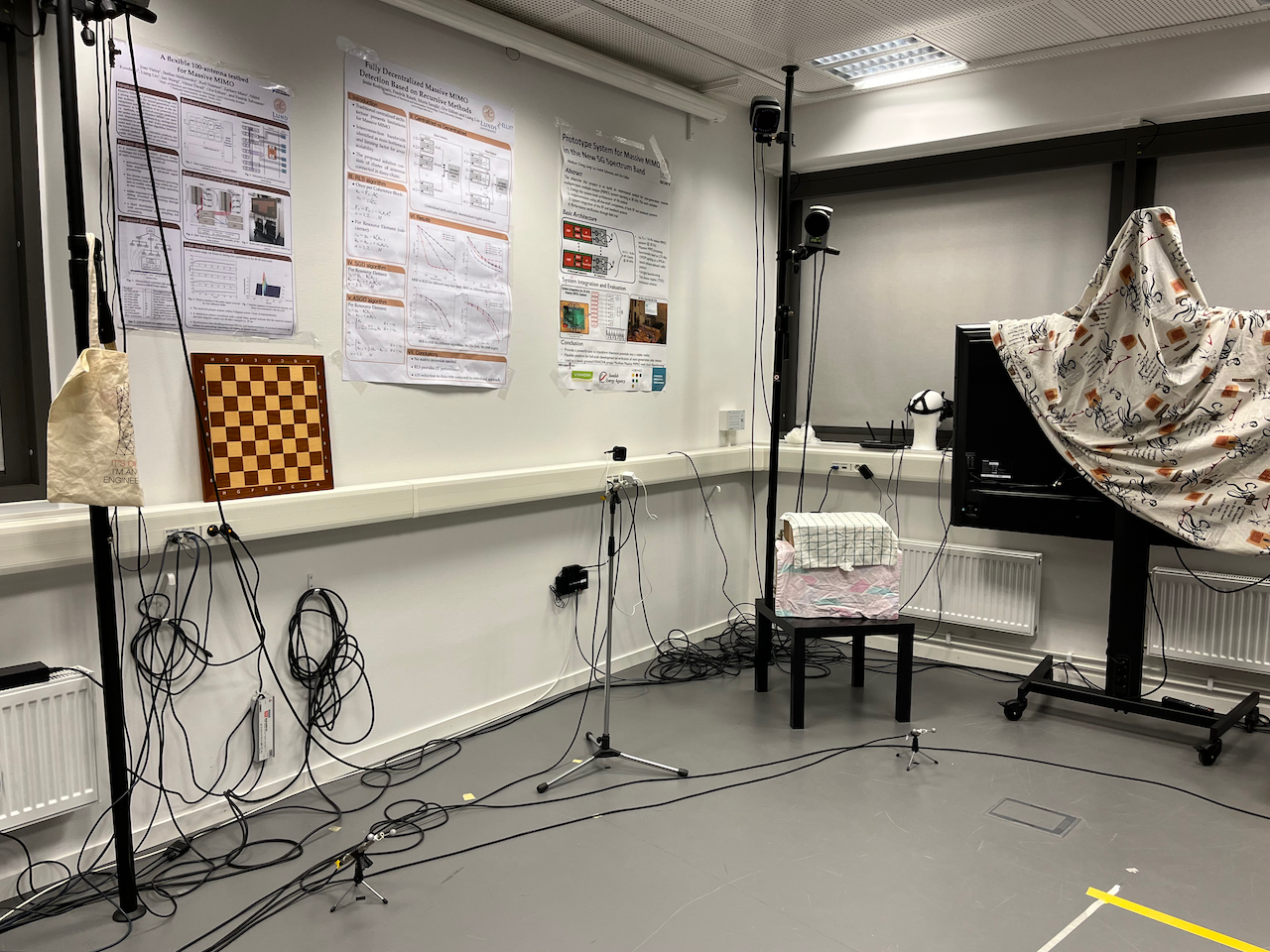}}}
  \caption{Examples of the decorations in the environment that can be used in vision-based localization algorithms.}
  \label{fig:Decorations} 
\end{figure}

While using these data, the effect of the rolling shutter camera can be ignored, since the camera is placed on top of a relatively stable robot moving at a low speed. The left and right imagers embed depth information to aid the generation of depth maps (with the help of a filter), and using these images in stereo-based localization algorithms might give lower accuracy compared to non-filtered images. The Mocap Studio is decorated with posters and objects with different textures to enable feature extraction in vision-based localization algorithms. Some example decorations can be seen in Fig.~\ref{fig:Decorations}.

\subsection{5G Radio system}
\begin{figure}[tb!]
  \centering
  \includegraphics[width=1.0\linewidth]{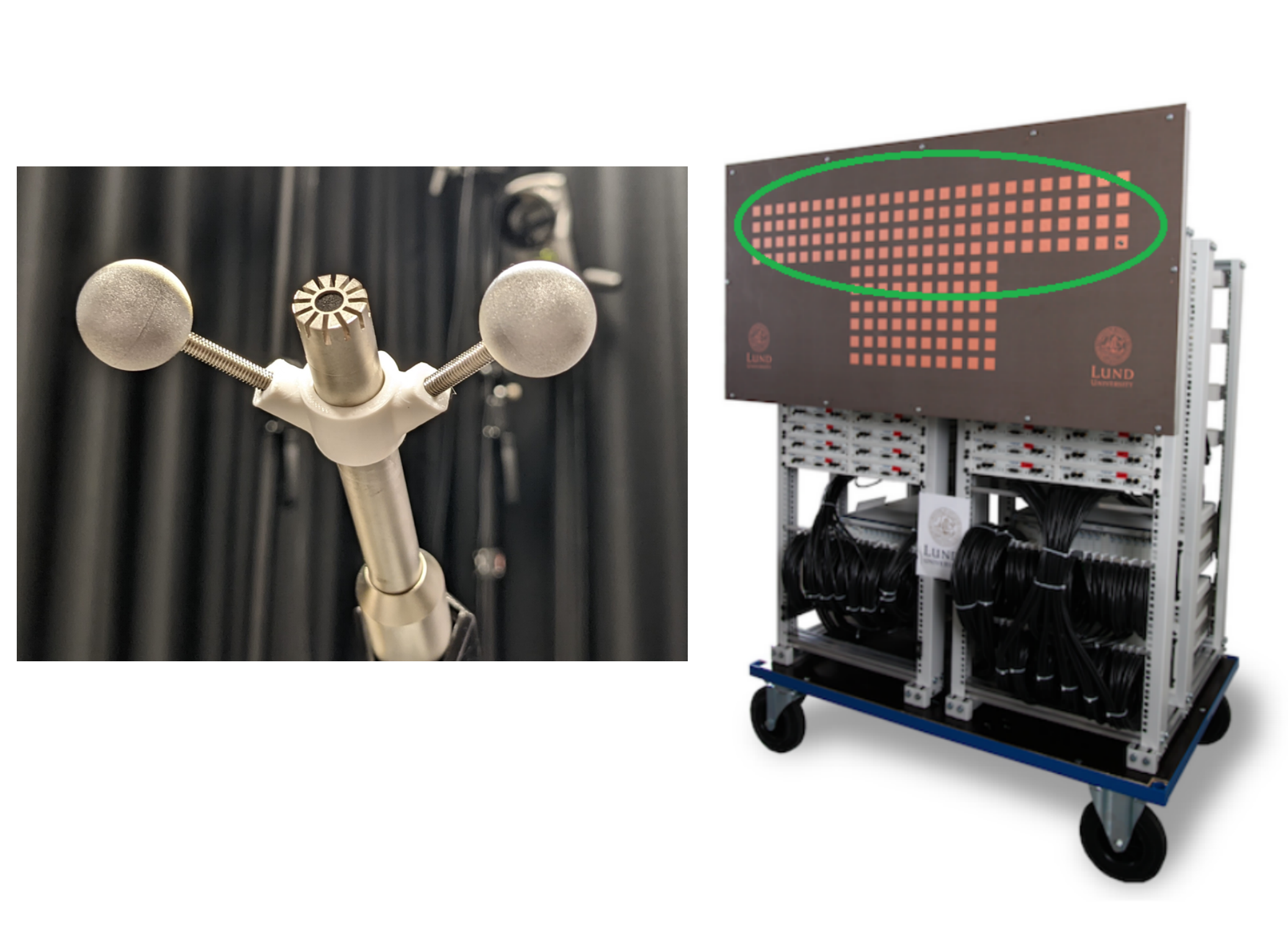}
  \caption{Markers and the microphone are shown on the left side of the image and the LuMaMi testbed where the wide antenna configuration is marked with the green circle is on the right.}
  \label{fig:LuMaMi_mic}
\end{figure}

The radio system consists of an user equipment (UE) which is connected to the single antenna mounted on the robot and the \textit{LuMaMi} testbed with $100$ antennas (each connected to an individual RF chain). \textit{LuMaMi} operates at a center frequency of $3.7$\,GHz and has a bandwidth of $20$\,MHz. The UE and \textit{LuMaMi} are connected via a cable to establish frequency synchronization and to eliminate carrier frequency offset. A collection of known signals, called pilot signals, are sent by the UE every $10$\,ms. The \textit{LuMaMi} testbed then captures $100$ snapshots of the channel per second (ps). These parameters are recorded and can be used to estimate the position of the mobile antenna in the environment. To exploit more information from the \textit{azimuth} compared to the \textit{elevation} domain, a wide antenna configuration, which is  $25 \times 4$, is selected rather than a quadratic one. The \textit{LuMaMi} testbed and the wide antenna configuration are depicted on the right side of Fig.~\ref{fig:LuMaMi_mic}. During the measurement campaign, the UE is placed on top of the moving robot, while \textit{LuMaMi} remains static. A universal software radio peripheral (NI X310) connected to a single dipole antenna is used as the UE.

\subsection{Audio system}
Audio-based localization uses an array of microphones to calculate the direction or location of a sound source by measuring different metrics, such as the time differences of arrival (TDOA) of sound signals at different microphones and triangulating the source's position based on these TDOA measurements. In our measurement campaign, $12$ microphones (\textit{T-bone MM-1}) are set up according to Fig.~\ref{fig:main_view} and connected to a single sound card that is recording to a laptop. A stereo speaker is placed on top of the robot, as well as the $12$th microphone. 
The microphone on the robot is placed as close as possible to the speaker (sound source) and works as a reference to synchronize the speaker with the microphones. In addition to the $12$ audio tracks, a synchronization pulse from the ground truth system on start and stop is recorded as a $13$th track (``Sync"). To make calculations easier by viewing the sound source as a point source, only one side of the speaker is enabled (playing sound), and the head of the reference microphone is placed directly in front of the sound source. All microphones, except the one on the robot, have two markers placed, as seen on the left side of Fig.~\ref{fig:LuMaMi_mic}. 


The sampling frequency of the microphones is $96$\,kHz and, if required, the audio system can localize the speaker at the same frequency. 
Microphones are placed asymmetrically on the floor and on different heights  to avoid microphones being co-linear or co-planar, as this may cause degeneracies when solving for positions. The sound level of every microphone is checked individually and the speaker is tested. 

\section{Calibration and Synchronization}

In this section, we outline the calibration and synchronization procedures of different systems, which are key requirements for localization using each sensor and using the sensors together. Calibrating a sensor is important to ensure accurate and reliable measurements, while time synchronization is crucial for multisensory applications to ensure that data from different sensors are aligned in a common temporal reference timeframe, enabling accurate correlation and fusion of information. Moreover, the parameters obtained by the sensor calibration are usually required as input to the localization algorithms. 

\subsection{Intrinsic and Extrinsic Calibration of the Sensors}

As an overview, the camera (and the built-in IMU) used for the vision system is calibrated intrinsically and extrinsically while the intrinsic calibrations of the sensors used for radio and audio systems are not done by us. Thus, for the radio system, traditional radio-based localization methods cannot be used, since the corresponding amplitude and phase of the antennas are not calibrated.

For the camera, two intrinsic calibrations are performed. For the first calibration, we follow the calibration procedure proposed by the manufacturer for the best performance. The second calibration is done using the \textit{Kalibr} calibration tool~\cite{kalibr}. With this calibration, the intrinsic and extrinsic parameters of the camera are extracted. The intrinsic parameters obtained with this method include camera centers, focal length, distortion parameters, and noise information for the IMU. The relationship between different cameras (i.e. two imagers and the RGB camera) and the transformation matrix between the cameras and IMU are obtained and used as extrinsic parameters of the camera. 

Temperature changes can significantly affect the calibration of the sensors. The internal temperature change of the devices was not recorded, but the room temperature has been measured for some of the trajectories. The change in temperature also affects the propagation factors of the air and the speed of sound, which can have an impact on the audio localization algorithms. 



\subsection{Time Synchronization}

\begin{figure}[tb!]
  \centering
  \includegraphics[width=1.0\linewidth]{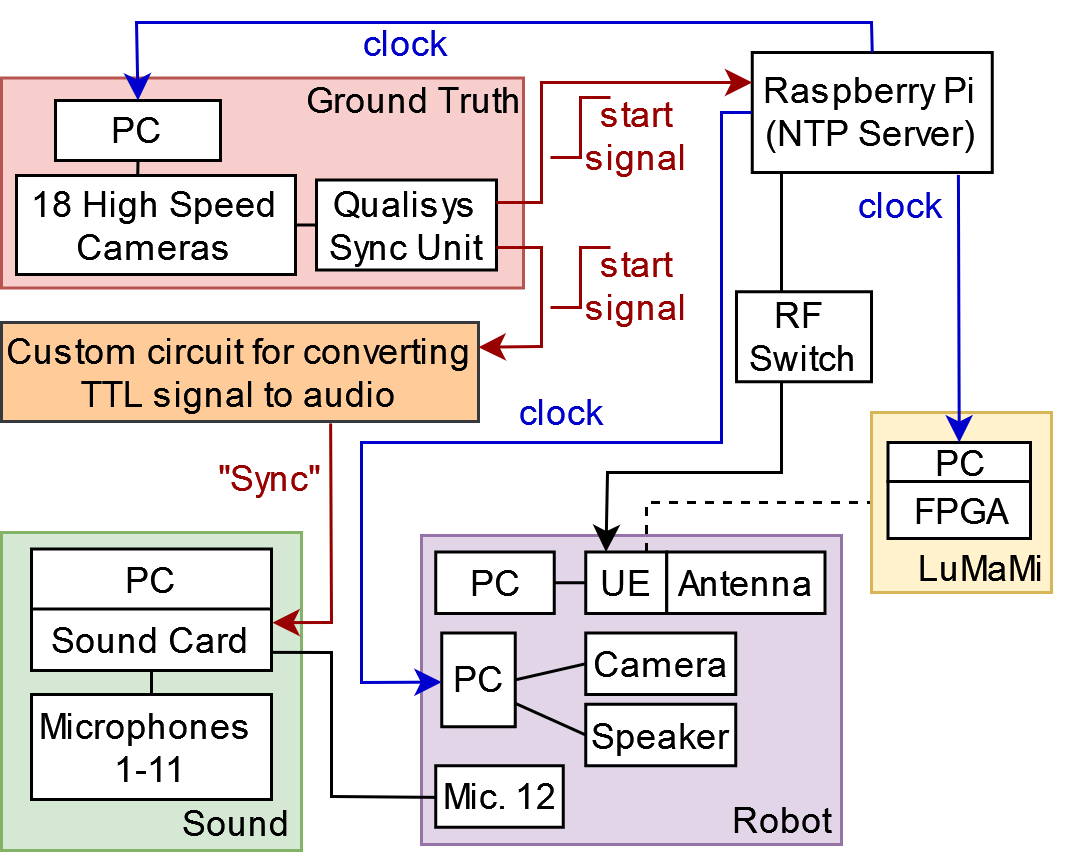}
  \caption{The overview block diagram of the synchronization system and the connections to different sensors.}
  \label{fig:MC_Sync_timing}
\end{figure}

We take several steps to make sure that the timestamps can be matched for each sensor (and the ground truth), so that the collected data can be verified with the ground truth data and used for sensor fusion. The block diagram of different systems and the connections that enable the time synchronization are given in Fig.~\ref{fig:MC_Sync_timing}. The time synchronization relies on the \textit{Raspberry Pi} and \textit{Qualisys Sync Unit}. An Network Time Protocol (NTP) server is used to give the correct timestamps to each system, even though there are other circuits to sync the different systems in some cases. 

The NTP server runs on the \textit{Raspberry Pi} and used to synchronize all computers regularly, except the computer used for sound recordings. One of the \textit{Raspberry Pi} Input/Output pins, configured as input with interrupt function, is connected to the \textit{Qualisys Sync Unit}, in order to listen to the short pulse (TTL signal) sent by the ground truth system when the recording starts. This TTL signal is called the ``start signal" in Fig.~\ref{fig:MC_Sync_timing} and is also sent to the sound system, connecting all the systems together. 
A more detailed description of the synchronization methods and the corresponding verification process are as follows. 

1) Vision: The NTP server on the \textit{Raspberry Pi} is used to update the clock in the laptop, which is used to record all the vision data. The accuracy of the system is validated up to $33$\,ms (the frame rate of the camera) with the help of an LED connected to the circuit used for the radio system. 

2) Radio: The UE and \textit{LuMaMi} are synchronized using a cable to avoid losing the signal between the receiver and the transmitter. The radio data is recorded on the testbed, thus its computer is connected to the NTP server to get the correct timestamp. To validate the timestamps, the \textit{Raspberry Pi} is connected to an RF switch on the robot, which can turn the transmission of the UE on and off. Afterwards, we plot when the UE starts and stops transmissions and compare the results to the timestamps obtained by the NTP server. 
The timestamp mismatch results in a maximum ground truth error $0.5$\,mm, which is negligible.

3) Audio: Synchronization between microphones is done by connecting all microphones to a single sound card. In order to synchronize the common timestamp, a separate circuit is built to convert the TTL signal from the synchronization box to an audio signal which is passed to the sound card as a separate channel, as described above. For the audio system, the $13$th ``Sync” channel is the recording of the pulse, which gets triggered when the mocap system starts and stops recording. The synchronization is verified using a clapperboard. The markers on the clapperboard are tracked by the ground truth system and seen by the vision system. When its two sections are clapped shut, it creates a distinctive sound and simultaneously marks a visual cue. 

\section{Created Trajectories}

There are $89$ trajectories collected in this measurement campaign that include all sensors. To meet all the requirements of different systems, we divide the dataset into two parts: the ``grid" and ``random" data. The ``grid" data consist of $75$ trajectories while ``random" data consist of $14$ trajectories. All the trajectories are divided into small sub-trajectories due to the limited amount of continuous data that \textit{LuMaMi} can store at a time due to the limited memory size. Based on previous experiments, we calculate the maximum amount of time that \textit{LuMaMi} can capture data continuously for up to $50$\,s (which includes the time it takes to establish the synchronization between all the systems). The system setup is the same in all trajectories, and if longer trajectories are required, the data can be used jointly instead. 

There are additional trajectories provided with the dataset that include data from one sensor at a time which can aid in calibration, extracting more information, or verifying the sensors separately. These include recordings with the microphones where the speaker is hand-held and the rest of the equipment is turned off, and recordings of the background noise in the environment when all the equipment is running. There are also recordings with the camera, where the camera is hand-held to cover the whole 3D environment, and a chessboard is recorded with different rotations and distances. 

\subsection{Grid Data}
One of the main considerations in designing trajectories is supporting dense sampling for radio-based localization algorithms. Thus, the spatial sampling of the data should be less than $\lambda/2$ (half a wavelength) according to the Nyquist theorem. For more details on this requirement, we refer the reader to~\cite{ICC}. The wavelength of \textit{LuMaMi} is calculated as approximately $8$\,cm, and as a result, training data is created such that it consists of measurements of the channel response approximately every $4$\,cm. In order to generate data that satisfies this condition, we scan the target area in the studio in $75$ trajectories and call this part the ``grid" data. 

\begin{figure}[tb!]
  \centering
  \includegraphics[width=0.90\linewidth]{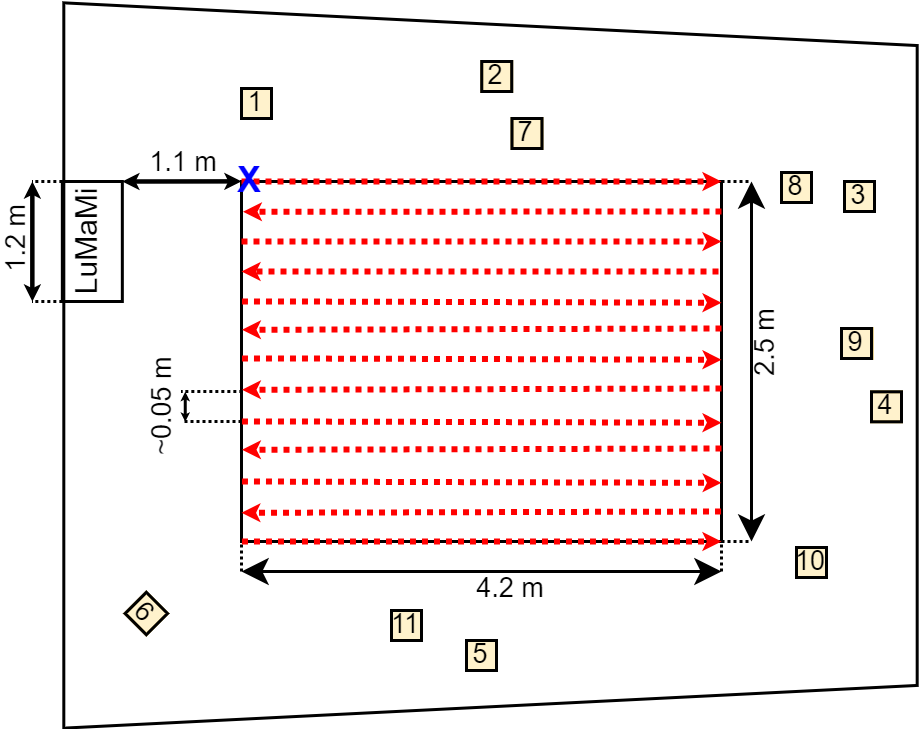}
  \caption{Red lines depicting the movement of the robot for the ``grid" trajectories where the blue cross is the initial location. Yellow numbered squares are the microphones.}
  \label{fig:grid_traj}
\end{figure}

To achieve the given resolution, the robot starts in the top-left corner of the effective area. With the help of ROS, the robot is moved $4.2$\,m in a straight line with a speed of $0.1 \text{\,m} \text{\,s}^{-1} $. Fig.~\ref{fig:grid_traj} shows the overall summary of the environment and the planned trajectory for this part. However, in the actual measurement campaign, a slight deviation from the straight line is seen for each ``grid" trajectory. As a result of changes in the trajectories and the limited time available in the Mocap Studio, the $4.2 \times 2.5 \textrm{\,m}^{2} $ area is covered with $75$ parallel trajectories for the ``grid" data. Throughout the ``grid" data, the speaker plays a chirp-based audio. 

\subsection{Random Data}

\begin{figure}[tb!]
  \centering
  \includegraphics[width=1.1\linewidth,trim={45cm 0 5cm 1cm},clip]{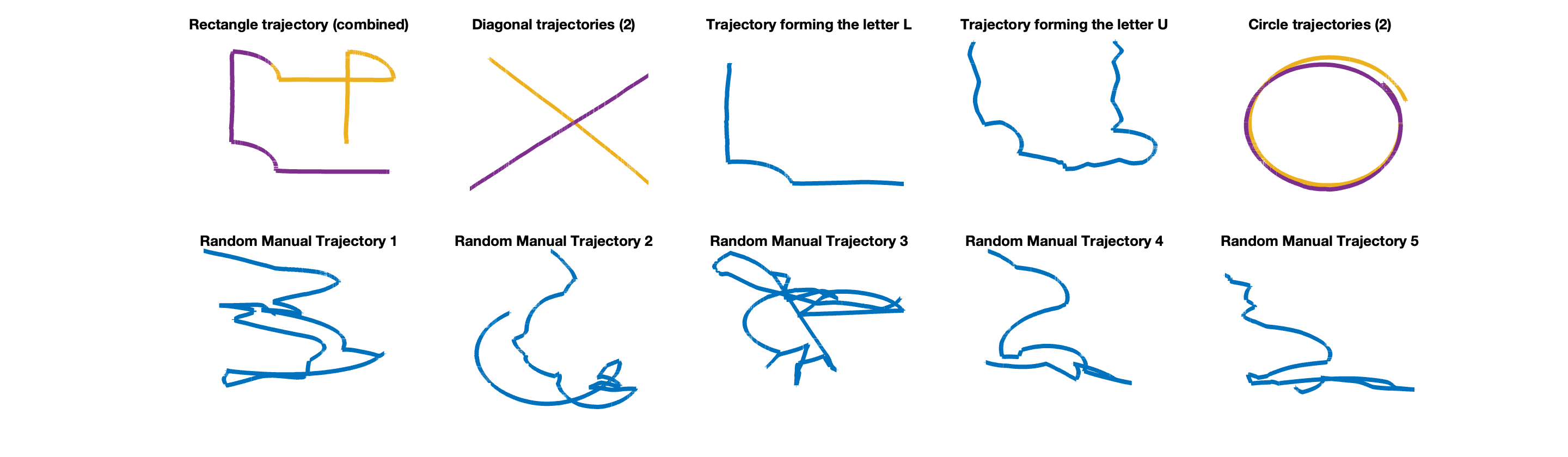} 
  \caption{Example ``random" trajectories plotted by using the data from the ground truth system tracking the camera object. Blue lines show one trajectory at a time where yellow and purple lines are used to indicate more than one trajectory.}
  \label{fig:random_trajectories}
\end{figure}

We conducted a total of $14$ measurements, which are called ``random" data, and some of the selected ones are shown in Fig.~\ref{fig:random_trajectories}. In this scenario, the robot orientation is not always perpendicular to the antenna array (unlike the ``grid" data), while humans act as static or dynamic scatterers in several trajectories. Dynamic environments pose significant localization challenges, including uncertainty and increased computational complexity, with numerous research works addressing these issues~\cite{human_influence, vision_dynamic_env}. 

The two circular trajectories are specifically designed for localization algorithms where loop closure is desirable. The two diagonal trajectories are corner cases for audio-based localization algorithms since they can cause some self-calibration methods for the audio system to fail~\cite{audio_problem}. In the manual trajectories, the robot is controlled by the built-in user interface with the virtual joystick. Thus, the speed of the robot is not constant, and the movements are more unstable and faster compared to the speed and movement of the robot in ``grid" data. The ``random manual" trajectories number 1, 2, and 6 include people moving around, whereas numbers 3, 4, and 5 do not. For the trajectories with people moving around in them, ground truth drops more frames due to markers not being seen by enough cameras. All the trajectories are in line-of-sight (LoS) from the point of view of \textit{LuMaMi} except for the ``random manual" trajectories with people walking around. In some cases in these trajectories, the LoS between \textit{LuMaMi} and the UE is obstructed by people and/or cables. We expect that the dynamic environment in these trajectories affects all localization methods, but these trajectories are particularly hard for radio-based localization algorithms due to the very dynamic environment. 

\section{Validation Experiments}

\begin{figure}[tb!]
  \centering
  \includegraphics[width=1.1\linewidth, trim={4.5cm 12cm 4cm 12cm},clip]{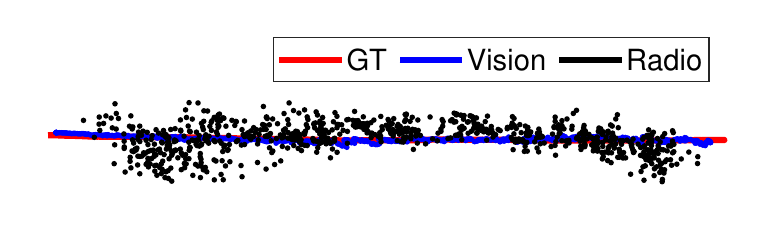} 
  \caption{The localization result of the ground truth (GT), vision, and radio based localization algorithms for \textit{Grid2} trajectory (single antenna is aligned to camera location).}
  \label{fig:radio_vision}
\end{figure}

This section offers a brief overview of the validation experiments done on the dataset to set a baseline for each component. For a more comprehensive understanding, we refer the interested reader to~\cite{yaman2023indoor}. In order to evaluate the dataset, we use RGB-D Oriented FAST and Rotated BRIEF simultaneous localization and mapping (ORB-SLAM3)~\cite{ORB-SLAM3} for the vision data, machine learning~\cite{ICC} for the massive MIMO radio data, and Structure from Sound2 (SFS2)\cite{9616051} for the audio data, respectively. Here, we use the \textit{Grid2} trajectory as an example. As shown in Fig.~\ref{fig:radio_vision}, the ORB-SLAM3 algorithm demonstrates a mean localization accuracy of $4.3$\,cm with a standard deviation (SD) of $2.2$\,cm for the \textit{Grid2} trajectory. 
The ground truth trajectory and the estimated trajectory are aligned by 
using the automated tools provided by~\cite{TUM_dataset}. The machine learning algorithm for the radio data is trained on the oddly numbered ``grid" trajectories. Evenly-numbered ``grid" trajectories are used to test the accuracy of the machine learning algorithm, which results in a mean error of $13$\,cm and SD of $12$\,cm. The SFS2 algorithm results in more than $1$\,m localization error in the ``grid" data where a chirp-based audio is used. In trajectories with music-based audio such as the \textit{Random Circle 2} trajectory, the current implementation of the algorithm can achieve up to a mean error of $9.6$\,cm and SD of $3.1$\,cm. Detailed evaluation and discussion of these results are not in the scope of this paper but can be found in~\cite{yaman2023indoor}. 

\section{Known Limitations and Remarks}

The following limitations and remarks should be considered when using the dataset: 

1) Due to the need for power supply and synchronization, a bundle of cables is connected to the robot. For some of the trajectories, the cables are hanging from the roof to avoid complications in the scene due to the robot's movement. For the radio system, the cables become a part of the channel, which might affect the accuracy of the localization algorithm. For the vision system, the cables are seen by the camera in some frames and create a dynamic environment. Moreover, the movement of the robot is restricted by the cables that are attached to it. 

2) In the ``grid" data, the robot is facing the same direction of the studio due to the previous remark/limitation (cables). This results in the same viewing direction for the camera and the same direction between the antenna that is connected to the UE and \textit{LuMaMi}. For vision data, ``random" data can be used instead. 

3) The room temperature increases significantly throughout the measurements, due to the equipment in the studio. After the change in temperature is noticed, the studio is measured almost every hour for the second half of the ``grid" data. Unfortunately, no temperature data exist for the first half of the ``grid" data, where the studio got significantly warmer in the evening. We believe that the temperature was around \SI{28}{\celsius}. This is only important for the ``grid" data, and no significant temperature change was observed for the ``random" data.  
\section{Conclusion} 
Localization is an essential process for many different tasks and devices such as UAVs, autonomous cars, and service robots. Localization algorithms require high computational power, where low power and high efficiency are the key to enabling these new technologies in battery-operated devices. 
In this paper, we have described a novel, unique, and comprehensive dataset that includes synchronized multisensory data (vision, radio, and audio sensors), collected in a controlled environment, together with high precision ground truth labels. 
This dataset, fused or separately, should allow accurate localization of devices (within cm) in real time. 
We make the dataset publicly available and give detailed information regarding its creation, drawbacks, and limitations. Overall, we hope that this dataset will contribute to the creation of localization algorithms that maximize the usage of the available data for low-power devices and also provide an in-depth understanding and further development of AI, machine learning, and 5G (and beyond) applications.    

\section*{Acknowledgment}
The authors would like to thank the Humanities Lab, Volker Krueger, and Matthias Mayr from the Department of Computer Science, Anders Robertsson from the Department of Automatic Control, and Sirvan Abdollah Poor, Jesús Rodríguez Sánchez, Jens Gulin and Sara Gunnarsson from the Department of Electrical and Information Technology at Lund University for providing resources, technical support, assistance, and feedback in our measurement campaign. In Memoriam of Anders Robertsson. 

\newpage
\bibliographystyle{ieeetr}
\bibliography{main}

\end{document}